\documentclass[11pt,a4paper,twoside]{article}
\usepackage{amsmath}
\usepackage{amsfonts}
\usepackage{changes}

\usepackage{color}
\usepackage{graphicx}
\usepackage{lmodern}
\usepackage[T1]{fontenc}
\usepackage[utf8]{inputenc} 
\numberwithin{equation}{section}
\usepackage{comment}
\usepackage{tabularx}
\usepackage{tikz}
\usepackage{pgfplots}
\usepackage{mathtools}

\makeatletter
\newcounter {subsubsubsection}[subsubsection]
\renewcommand\thesubsubsubsection{\thesubsubsection .\@arabic
	\c@subsubsubsection}
\newcommand\subsubsubsection{\@startsection{subsubsubsection}{4}{\z@}%
	{-3.25ex\@plus -1ex \@minus -.2ex}%
	{1.5ex \@plus .2ex}%
	{\normalfont\normalsize\bfseries}}
\newcommand*\l@subsubsubsection{\@dottedtocline{4}{10.0em}{4.1em}}
\newcommand*{\subsubsubsectionmark}[1]{}
\makeatother

\usepackage{ifthen} 
\newboolean{pdflatex}
\setboolean{pdflatex}{true} 

\usepackage{ulem}
\usepackage{caption}
\usepackage{subcaption}
\usepackage{multirow}
\usepackage{amssymb}
\usepackage{longtable}
\usepackage{rotating}
\usepackage{graphicx}
\usepackage{booktabs}
\usepackage{xparse}
\NewDocumentCommand{\rot}{O{45} O{1em} m}{\makebox[#2][l]{\rotatebox{#1}{#3}}}%

\newboolean{articletitles}
\setboolean{articletitles}{true} 

\newboolean{uprightparticles}
\setboolean{uprightparticles}{false} 

\newboolean{inbibliography}
\setboolean{inbibliography}{false} 

\usepackage{longtable} 

\begin{document}

\noindent

{\bf
{\Large Constraints on interacting dynamical dark energy and a new test for $\Lambda$CDM

}
} 

\vspace{.5cm}
\hrule

\vspace{1cm}

\noindent

{\large\bf{Marco Bonici\footnote{\tt marco.bonici@ge.infn.it } and Nicola Maggiore\footnote{\tt nicola.maggiore@ge.infn.it }
\\[1cm]}}

\setcounter{footnote}{0}

\noindent
{{}Dipartimento di Fisica, Universit\`a di Genova,\\
via Dodecaneso 33, I-16146, Genova, Italy\\
and\\
{} I.N.F.N. - Sezione di Genova\\
\vspace{1cm}

\noindent
{\tt Abstract~:}

\vspace{.5cm}

We consider a generic description of {interacting} dynamical Dark Energy, characterized by an equation of state with a time dependent coefficient $w(t)$, and which may interact with both radiation and matter. Without referring to any particular cosmological model, we find a differential equation which must be satisfied by $w(t)$ and involving the function $Q(t)$ which describes the interaction between Dark Energy and the other cosmological fluids. The relation we find represents a constraint for various models of {interacting dynamical} Dark Energy. In addition, an observable is proposed, depending on kinematic variables and on density parameters, which may serve as a new test for $\Lambda$CDM.

\newpage

\section{Introduction}

The accelerating expansion of our Universe 
\cite{Riess:1998cb, Perlmutter:1998np} can be described by  a cosmological constant in the Einstein-Hilbert action of General Relativity,  introduced by Einstein \cite{Einstein:1917ce} in order to have a static Universe. There are a few well known good reasons to be unsatisfied with the description of Dark Energy (DE) in terms of a cosmological constant, but the known difficulties, or seemingly unnatural coincidences, can be solved invoking very peculiar initial  conditions \cite{Amendola:2016saw}. In the hopeful wait of an experimental conclusive evidence, theorists since long time provided us with a variety of alternative models for DE \cite{Copeland:2006wr}, with the request that any cosmological model should reproduce  an Universe which, at our epoch, is almost perfectly flat and filled by matter and DE in the ratio of about 3/7, where the DE is effectively approximated by a constant.  {In Literature many ``dynamical'' alternatives for DE can be found, like, for instance,  the quintessence models \cite{Ratra:1987rm,Wetterich:1987fm}, where the role of the cosmological constant is  played by scalar potentials, suitably parametrized to get the desired behavior, and the K-essence models \cite{Chiba:1999ka,ArmendarizPicon:2000dh,ArmendarizPicon:2000ah}, likewise built in terms of scalar fields, where the accelerated expansion of the Universe is driven by the kinetic term.} Both quintessence and K-essence models belong to the wider category of modified theories of gravity, whose purpose is to extend their range of validity to large, galactic, scales. In the most general case, any dynamical, as opposed to constant, model for DE may interact with all the components of the cosmological perfect fluid in terms of which is written the energy momentum tensor appearing at the right hand side of the Einstein equations
\begin{equation}
R_{\mu\nu}-\frac{1}{2}g_{\mu\nu}R=8\pi G\ T_{\mu\nu}\ .
\label{0.1}\end{equation}
 The coupling could be minimal, through the metric dependent invariant measure, {or} non-minimal, with direct and non-trivial coupling with gravity, like in the scalar-tensor theories, or by means of direct interactions with baryonic matter, and/or with neutrinos, and/or with Dark Matter 
\cite{Ellis:1989as,Damour:1993id,Damour:1994zq,Amendola:1999er,Amendola:1999qq,Szydlowski:2005ph}.

In this paper we keep a very general perspective. Without referring to any particular DE model, we assume only that DE is realized by means of a perfect fluid satisfying an Equation of State (EoS) with a time dependent $w$-coefficient 
\begin{equation}
p_{\mbox{\tiny DE}}=w(t)\rho_{\mbox{\tiny DE}}
\label{0.2}\end{equation} 
and that DE interacts non-minimally with any cosmological component. The interactions result in  broken covariant conservation laws of the energy momentum tensors {of} the single cosmological components
\begin{equation}
\nabla_\mu (T_i)^\mu_\nu=(Q_i)_\nu\ \ \mbox{$i$=matter, radiation, DE}
\label{0.3}\end{equation}
keeping the total energy momentum tensor  conserved. The aim of this paper is to give a criterion to select amongst different models of {interacting} dynamical DE, assuming only the validity of the Friedmann equations for the scale factor appearing in the Robertson-Walker metric. 
This subject has been faced following different strategies \cite{Wang:2016lxa,Bolotin:2013jpa,Yang:2018qec,Yang:2018uae,Chimento:2003iea,Farrar:2003uw,Guo:2007zk,Yang:2018euj,vonMarttens:2018iav,Aviles:2012ay}, all of which need some kind of assumptions, on the phenomenological form of the interactions $Q_i$, or on the choice of the potential in the quintessence models, for instance. In this paper, we try to be as much general as possible, adopting a model independent cosmographic approach (see \cite{Bolotin:2018xtq} for an updated review).

In order to reach this goal, in Section 2 we relate the time derivative of the DE EoS coefficient $w(t)$ to the interactions $Q(t)$, by means of the kinematic variables associated to the the scale factor $a(t)$: the Hubble parameter $H(t)$, the deceleration $q(t)$ and the jerk $j(t)$, {sometimes called statefinder \mbox{$r(t)$-parameter \cite{Sahni:2002fz}}}. In Section 3 we discuss the implications of our analysis for the $\Lambda$CDM model and we propose an observable, written in terms of kinematic variables and density parameters, whose non-vanishing value would imply a failure of $\Lambda$CDM.
In the concluding Section 4 we summarize and discuss our results.

\section{Constraints {on} interacting dynamical Dark Energy}

The energy momentum tensor for a cosmological perfect fluid is: 
\begin{equation}
T_{\mu\nu}=(\rho+p)U_\mu U_\nu+pg_{\mu\nu}\ ,
\label{1}\end{equation}
where $U_\mu$ is the fluid four-velocity, $\rho$ is the rest-frame energy density and $p$ is the isotropic rest-frame pressure. The EoS relates pressure and energy density and its general form is: 
\begin{equation}
p=p(\rho)\ ,
\label{2}\end{equation}
whose simplest case is represented by the linear relation
\begin{equation}
p=w\rho\ ,
\label{3}\end{equation}
where $w$ is a coefficient not depending from the energy density $\rho$.

Following \cite{Visser:2003vq}, we consider here the more general EoS \eqref{2}, whose
Taylor expansion around the energy density at the present epoch $\rho_0=\left.\rho(t)\right|_{t=t_0}$ is
\cite{Visser:2003vq}
\begin{equation}
p(\rho)=p_0 + \kappa_0 (\rho-\rho_0) + {\cal O}[(\rho-\rho_0)^2]\ ,
\label{4}\end{equation}
where $p_0=p(\rho_0)$ and $\kappa_0=\left.\frac{dp}{d\rho}\right|_{t=t_0}$\footnote{From now on, 
${\cal O}_0\equiv\left.{\cal O}(t)\right|_{t=t_0}$, {for any observable ${\cal O}(t)$}, where $t_0$ stands for the present epoch.}.

The aim is to express the first two coefficients of the above expansion in terms of the scale factor $a(t)$ appearing in the Robertson - Walker metric
\begin{equation}
ds^2=-dt^2+a^2(t)\left[\frac{dr^2}{1-kr^2}+r^2(d\theta^2+\sin^2\theta d\phi^2)\right]\ ,
\label{4.1}\end{equation}
where $k$ is a constant parameter related to the spatial curvature: $k=0,  k>0$ and $k<0$ for flat, closed and open Universes, respectively.

More precisely, we would like to write $p_0$ and $\kappa_0$ in terms of the kinematic variables related to the dimensionless time derivatives of $a(t)$, namely the Hubble parameter $H(t)$, the deceleration $q(t)$ and the jerk $j(t)$, which are observables quantities, thus making this approach independent from a particular cosmological model. The kinematic variables are  defined as follows
\begin{eqnarray}
H &=& \frac{\dot{a}}{a} \label{5} \\
q &=& -\frac{\ddot{a}}{aH^2} \label{6} \\
j &=& \frac{\dddot{a}}{aH^3}\label{7}\ .
\end{eqnarray}

It is customary to suppose that the cosmological fluid is an incoherent mixture of the three forms of canonical fluids ($i=1$ matter, $i=2$ radiation, $i=3$ DE represented by a cosmological constant $\Lambda$) plus, following a standard notation \cite{Carroll:2004st},  the spatial curvature contribution ($i=4$ ), each satisfying the linear EoS
\begin{equation}
p_i=w_i\rho_i\ .
\label{8}\end{equation}
%
%
Consequently, the EoS \eqref{3} takes the form 
\begin{equation}
\sum_{i=1}^4p_i=\sum_{i=1}^4w_i\rho_i\ .
\label{9}\end{equation}
Once we define the density parameters
\begin{equation}
\Omega_i=\frac{8\pi G}{3H^2}\rho_i\ ,
\label{10}\end{equation}
the Friedmann equation can be written
\begin{equation}
\sum_i\Omega_i=1\ .
\label{11}\end{equation}
As done in \cite{Visser:2003vq}, we denote with $\overline{\cal O}$ the  average of {generic physical observables} ${\cal O}_i$ weighted by the density parameters $\Omega_i$ of each fluid \cite{Visser:2003vq}: 
\begin{equation}
\overline{\cal O} \equiv \sum_i^4{\cal O}_i\Omega_i\ .
\label{12}\end{equation}
Consequently, from the EoS \eqref{3} we have
\begin{equation}
w=\frac{p}{\rho}=\frac{\sum_ip_i}{\sum_i\rho_i}=\frac{\sum_iw_i\rho_i}{\sum_i\rho_i}=
\frac{\sum_iw_i\Omega_i}{\sum_i\Omega_i}=\sum_iw_i\Omega_i=\overline{w}\ .
\label{13}\end{equation}

In this paper we introduce the following two generalizations with respect to the approach described in \cite{Visser:2003vq}:
\begin{enumerate}
\item We allow a time dependence of the EoS coefficients $w_i$ appearing in \eqref{8}
\begin{equation}
p_i=w_i(t)\rho_i\ .
\label{14}\end{equation}
Even though we are mostly interested in physical situations where only the DE fluid may have a time dependent $w_3(t)$, for the moment we take a more general attitude. The known scalar quintessence model for DE is an example of DE fluid with a time dependent EoS coefficient, but we point up that in this paper we do not necessarily limit ourselves to this particular case.
\item
The energy momentum tensor \eqref{1} is given by the sum of the different components of the perfect cosmological fluid. We give the possibility to each component to break the conservation law: 
\begin{equation}
\nabla_\mu (T_i)^\mu_{\ \nu}=(Q_i)_\nu\ ,
\label{15}\end{equation}
keeping the total energy momentum tensor conserved, which implies a constraint on the  breakings
\begin{equation}
\nabla_\mu T^\mu_{\ \nu}=0\Rightarrow\sum_i(Q_i)_\nu=0\ .
\label{16}\end{equation}
In most cases, only the matter and DE components of the energy momentum tensor possibly display a breaking of the conservation law in the late Universe,  not the radiation nor the curvature contributions. Again, for the moment we stay on general grounds, and the breakings $Q_i$, which, because of the constraint \eqref{16} must be at least two, physically correspond to interactions between the cosmological components fluids.
Examples of non-vanishing DE interactions are given in \cite{Wang:2016lxa,Bolotin:2013jpa,Yang:2018qec,Yang:2018uae,Chimento:2003iea,Farrar:2003uw,Guo:2007zk,Yang:2018euj}.
\end{enumerate}

Deriving both sides of the Friedmann equation \eqref{11} with respect to time, we have 
\begin{equation}
\sum_i\dot\Omega_i=0\Rightarrow\sum_i \frac{d}{dt}\left(\frac{\rho_i}{H^2}\right)=0\ .
\label{17.1}\end{equation}
To calculate $\dot\rho_i$, we use the covariant conservation of the energy momentum tensor \eqref{1}, 
The $\nu=0$ component of \eqref{15} gives
\begin{equation}
\dot\rho_i=-3H(\rho_i+p_i)+Q_i\ ,
\label{17}\end{equation}
where we defined
\begin{equation}
Q_i\equiv-(Q_i)_0=+(Q_i)^0\ .
\label{18}\end{equation}

On the other hand
\begin{equation}
\dot{H}=\frac{\ddot{a}}{a}-\left(\frac{\dot{a}}{a}\right)^2=-H^2(1+q)\ ,
\label{19}\end{equation}
where we used the definition \eqref{6} of the deceleration $q(t)$.

Coming back to Eq. \eqref{17.1}, we can now write
\begin{eqnarray}
0 
&=&
\sum_i\left(\frac{\dot\rho_i}{H^2}-\frac{2}{H^3}\rho_i\dot{H}\right) \label{20} \\
&=&
\sum_i\left[
\frac{-3H(1+w_i)\rho_i}{H^2}
+\frac{Q_i}{H^2}
+ \frac{2}{H^3}\rho_iH^2(1+q)\right] \label{21} \\
&=&
-3H(1+\overline{w})+2H(1+q)\ ,
\label{22}
\end{eqnarray}
where we used \eqref{17}, \eqref{8} and \eqref{19}, and we used the definition of weighted average \eqref{12} for the quantities $w_i(t)$ and the constraint \eqref{16} on the breakings $Q_i$.

Therefore, using \eqref{13}, the following relation holds
\begin{equation}
\overline{w}=w=\frac{2q-1}{3}\ ,
\label{23}\end{equation}
which, in particular, relates the first coefficient of the Taylor expansion of the EoS  \eqref{4} to the deceleration $q(t)$, since
\begin{equation}
p_0=w_0\rho_0=\frac{2q_0-1}{3}\rho_0\ .
\label{24}\end{equation}
Notice that the relation \eqref{23}, which has been derived in \cite{Visser:2003vq} for non-interacting DE with constant EoS $w$-parameter,  has a general validity, since it holds also for
$\dot{w}_i\neq 0$ and $Q_i\neq 0$.

Let us now consider $\kappa_0$, the second coefficient of the EoS Taylor expansion \eqref{4}: 
\begin{eqnarray}
\kappa
&=&
\frac{dp}{d\rho}
=
\frac{\dot{p}}{\dot{\rho}}
=
\frac{\sum_i\dot{p}_i}{\sum_i\dot{\rho}_i}
=
\frac{\sum_i(w_i\dot\rho_i+\dot{w}_i\rho_i)}{\sum_i\dot{\rho}_i} \nonumber\\
&=&
\frac{\sum_i\{w_i[-3H(1+w_i)\rho_i+Q_i]+\dot{w}_i\rho_i\}}{\sum_i[(-3H)(1+w_i)\rho_i+Q_i]} \nonumber\\
&=&
\frac{\sum_i [ 
(-3H)(w_i+w^2_i)\rho_i + w_iQ_i + \dot{w}_i\rho_i]}
{\sum_i(-3H)(1+w_i)\rho_i} \nonumber\\
&=&
\frac{\overline{w}+\overline{w^2}}{1+\overline{w}}
- \frac{8\pi G}{9H^3} \frac{\sum_iw_iQ_i}{1+\overline{w}}
-\frac{\overline{\dot{w}}}{3H(1+\overline{w})}\label{25}\ ,
\end{eqnarray}
where we took into account \eqref{17}, \eqref{16} and \eqref{12}. We need $\overline{\dot{w}}$, $i.e.$ the weighted average of the time derivatives of the EoS coefficients $w_i(t)$, which vanish in the $\Lambda$CDM model.  To obtain it, we look for an expression for the time derivative of the weighted average $\dot{\overline{w}}$: 
\begin{eqnarray}
\dot{\overline{w}}
&=&
\sum_i \frac{d}{dt}(w_i\Omega_i)
=
\overline{\dot{w}} 
+ \frac{8\pi G}{3}\sum_iw_i \frac{d}{dt}\left(\frac{\rho_i}{H^2}\right)
\nonumber \\
&=&
\overline{\dot{w}} 
+\frac{8\pi G}{3} \sum_i\left(
\frac{\dot{\rho}_i}{H^2}-\rho_i\frac{2}{H^3}\dot{H}\right)
\nonumber \\
&=&
\overline{\dot{w}} 
+\frac{8\pi G}{3}
\sum_i \left[
\frac{(-3H)(1+w_i)\rho_i+Q_i}{H^2}
+\rho_i\frac{2}{H}(1+q)\right]
\nonumber \\
&=&
\overline{\dot{w}} 
-3H(\overline{w}+\overline{w^2})+\frac{8\pi G}{3H^2}\sum_iw_iQ_i + 3H\overline{w}(1+\overline{w})\ ,
\label{26}\end{eqnarray}
where, in the last row, we used \eqref{23} to eliminate the deceleration $q$ in favor of $\overline{w}$.
Introducing the variance of the values $w_i$
\begin{equation}
\sigma^2_w=\overline{w^2}-\overline{w}^2\ ,
\label{27}\end{equation}
we get 
\begin{equation}
\overline{\dot{w}} =\dot{\overline{w}} + 3H\sigma^2_w-\frac{8\pi G}{3H^2}\sum_iw_iQ_i\ .
\label{28}\end{equation}
It is easily seen that, using in \eqref{25} the above expression \eqref{28} for $\overline{\dot{w}}$ and the definitions of the deceleration $q(t)$ \eqref{6} and of the jerk $j(t)$ \eqref{7}, we finally get
\begin{equation}
\kappa = \frac{dp}{d\rho}=\frac{j-1}{3(1+q)}\ ,
\label{29}\end{equation}
which, as $w(t)$ \eqref{23}, is an universal quantity, whose expression is valid whether $\dot{w}_i\neq0$ and $Q_i\neq0$ or not.

Let us take for a moment Eq. \eqref{28} at $\dot{w}_i=Q_i=0$, which is the standard case we are generalizing in this paper. The variance \eqref{27} reduces to
\begin{equation}
\sigma_w^2=-\frac{\dot{\overline{w}}}{3H}=\frac{2}{9}[j-q(1+2q)]\ ,
\label{30}\end{equation}
where $\overline{w}(t)$ in \eqref{23}  and the definition \eqref{7} of the jerk $j(t)$ have been used. The above expression for $\sigma_w^2$ tells us how the weighted accuracy on the estimate of the $w_i$, assumed to be constant, evolves in time, driven by the time dependence of the cosmological parameters $\Omega_i$ only. In general, it is not allowed to deduce that the right hand side of \eqref{30} is non-negative, since the weights $\Omega_i$ present in 
$\sigma_w^2=\sum_i(w_i-\overline{w})^2\Omega_i$ may be negative. Indeed, while $\Omega_1$ and $\Omega_2$ are certainly non-negative functions of time, since they are related to matter and radiation energy density respectively, the density parameters $\Omega_3$ and $\Omega_4$, which refer to DE and curvature, might, in principle, have any sign. What we can state, is that, at our epoch, 
$\Omega_1^{(0)}\simeq 0.3$, $\Omega_2^{(0)}\simeq 0$ and $\Omega_3^{(0)}\simeq 0.7$ \cite{Riess:1998cb}, and, consequently, that the Universe, in excellent approximation,  is spatially flat $k\simeq 0$. Therefore, at our epoch, but not at any time, the right hand side of \eqref{30} is non-negative
\begin{equation}
j_0\geq q_0(1+2q_0)\ .
\label{32}\end{equation}
Eq. \eqref{32} is a constraint which must be satisfied, at $t=t_0$, by the kinematic variables related to the time derivatives of the scale factor $a_0$, namely the deceleration $q_0$ and the jerk $j_0$.

In case of non-vanishing $\dot{w}_i$ and $Q_i$, the more general relation \eqref{28} represents a constraint for interacting dynamical DE. In fact, it relates the possible time dependent $w_i(t)$ appearing in the EoS \eqref{8} of the cosmological fluids \eqref{3} to their corresponding interactions \eqref{15}. At the present epoch $t=t_0$ we have: 
\begin{equation}
\left. \dot{\overline{w}}+3H\sigma_w^2 \right|_{t=t_0} \equiv K_0
=
\left. \sum_i(\dot{w}_i\Omega_i + \frac{8\pi G}{3H^2}w_iQ_i) \right|_{t=t_0}\ ,
\label{33}\end{equation}
where $K_0$ is a physical observable, depending on measurable quantities (density parameters and kinematic variables).

Making the reasonable assumption that only the DE component of the cosmological perfect fluid may have an EoS of the form \eqref{8} with $\dot{w}_3(t)\neq0$, and observing that the relevant interactions are the ones involving DE, which translates into $Q_3\neq 0$, the relation \eqref{33} at $t=t_0$ reduces to
\begin{equation}
\left. \dot{w}_3\Omega_3+\frac{8\pi G}{3H^2}w_3Q_3\right|_{t=t_0}=K_0\ ,
\label{34}\end{equation}
where we used the fact that, for matter, $w_1$ strictly vanishes.

Since the values at our epoch  of the DE density parameter $\Omega_3^{(0)}$, of the coefficient of the DE EoS $w_3^{(0)}$, of the Hubble constant $H_0$ and of the quantity $K_0$, are known, the relation \eqref{34} represents a constraint on the possible theoretical models of {interacting dynamical} DE, in particular  {
on the time dependence of the DE EoS coefficient $\dot{w}_3^{(0)}$ and on the  DE interaction $Q_3^{(0)}$
\begin{eqnarray}
w_3(t) &=& w_3^{(0)} + \dot{w}_3^{(0)} (t-t_0) + O(t^2) \label{35}\\
Q_3(t) &=& Q_3^{(0)}  + O(t)\ . \label{36}
\end{eqnarray}
It is a remarkable and, to our knowledge, so far unknown fact that the interactions involving DE and its dynamical EoS are not independent one from each other.
}

\section{A new test for $\Lambda$CDM model}

The aim of this paper is to put constraints, mainly by means of the relation \eqref{34}, on the possible models of DE, with particular concern on the DE EoS and on the interaction DE-matter. In order to be able to make a comparison, it is useful to summarize what is predicted by the Standard Model of Cosmology.

After the observational evidence from supernovae for an accelerating Universe and a cosmological constant \cite{Riess:1998cb}, we know that, at our epoch, our Universe is filled by DE and (mostly dark) matter : 
\begin{equation}
\Omega^{(0)}_1\simeq 0.3\ ;\ \Omega^{(0)}_3\simeq 0.7\ ,
\label{37}\end{equation}
where we used the notations adopted in this paper, according to which the subscripts $1$ and $3$ stand for matter and DE, respectively. An immediate consequence of the Friedmann equation, is that our Universe is almost flat 
\begin{equation}
k\simeq 0\ ,
\label{38}\end{equation}
{
since, at our epoch, the radiation contribution to the whole cosmological perfect fluid being } 
is highly suppressed:
\begin{equation}
\Omega^{(0)}_2\simeq 0\ . 
\label{38.1}\end{equation}
The $\Lambda$CDM model  well describes this scenario, where the DE is realized by means of a cosmological constant $\Lambda$.  The $\Lambda$CDM situation, including the EoS coefficients $w_i$ of the single cosmological fluids, is summarized in the following Table 1: 
\begin{center}
\begin{tabular}{|l|c|c|c|c|}\hline
 & \footnotesize{i=1: matter} & \footnotesize{i=2: radiation} & \footnotesize{i=3: DE} & \footnotesize{i=4: curvature} 
\\ \hline
$\Omega^{(0)}_i$ &$0.3$ & $ 0$  & $ 0.7$ & $ 0$ 
\\ \hline
$w_i$ &0 &  1/3 & $-1$ & $-1/3$ 
\\ \hline
\end{tabular}

\vspace{.2cm}{\footnotesize {\bf Table 1.}
EoS coefficients and density parameters in $\Lambda$CDM.}
\end{center} 

In the $\Lambda$CDM  model the only EoS coefficient which survives is $w_3$. Its value ($w_3=-1$) corresponds to the contribution to the cosmological fluid coming from the cosmological constant.

According to the $\Lambda$CDM model, the jerk variable \eqref{7} should be constant, and in particular 
\begin{equation}
\Lambda\mbox{CDM}\ \ \Rightarrow\ \  j(t)=1\ .
\label{42}\end{equation}
This can be seen in many ways. In the particular framework of this paper, let us consider the expression \eqref{30} for the variance of the EoS coefficient $w$, which holds for $\dot{w}_i=Q_i=0$ and hence  true in the $\Lambda$CDM model:
\begin{equation}
j=\frac{9}{2}\sigma_w^2+q(2q+1)\ .
\label{43}\end{equation}
From the definition  of weighted average \eqref{12}, of variance \eqref{27}  and using the fact that in the $\Lambda$CDM model the only non-vanishing EoS coefficient is the DE one, we have
\begin{equation}
\sigma_w^2=w_3^2\Omega_3(1-\Omega_3)\ .
\label{44}\end{equation}
On the other hand, from $\overline{w}$ in \eqref{23}, we get the following relation for the deceleration parameter $q(t)$:
\begin{equation}
q = \frac{3}{2}w_3\Omega_3+\frac{1}{2}\ .
\label{45}\end{equation}
Using \eqref{44} and \eqref{45}, we get the following relation for the jerk parameter
\begin{equation}
j=1 + \frac{9}{2}w_3(1+w_3)\Omega_3\ ,
\label{46}\end{equation}
which is equal to one if the DE is described by a cosmological constant, $i.e.$ $w_3=-1$. Therefore, the jerk parameter is a  model independent,  kinematic observable valuable to test the $\Lambda$CDM model, since any deviation from $j=1$ is a signal of alternative descriptions. 

Let us now consider $K_0$ defined by the left hand side of \eqref{33}
\begin{equation}
K_0\equiv \left. \dot{\overline{w}}+3H\sigma_w^2 \right|_{t=t_0}\ .
\label{47}\end{equation}
According to the $\Lambda$CDM model, the DE has a constant EoS $w$-coefficient, and does not interact, hence, from the right hand side of \eqref{33}, $K_0$ should vanish identically: 
\begin{equation}
\Lambda\mbox{CDM}\ \ \Rightarrow\ \  K_0=0\ .
\label{48}\end{equation}
From its definition \eqref{47}, it is easy to check that $K_0$ can be written in terms of measurable quantities as follows: 
\begin{equation}
\frac{K_0}{H_0}=3\Omega_1(1-\Omega_1)-\frac{2}{3}\left[j_0-q_0(1+2q_0)\right]\ .
\label{49}\end{equation}
We point out that, analogously to the case concerning the jerk parameter $j(t)\neq 1$, a non-vanishing value for $K_0$ would be a certain signal of the failure of the $\Lambda$CDM model, and we remark that the case $K_0\neq0$ is  independent of $j_0\neq1$. It is easily seen, in fact, that observational situations are possible where $j_0=1$ and $K_0\neq 0$ at the same time, for which we should conclude against $\Lambda$CDM.

On the other hand, even a $K_0$ compatible with zero would not represent a confirmation of $\Lambda$CDM. Both cases $K_0=0$ and $K_0\neq 0$, in fact, could be realized by means of an {interacting} dynamical DE, with $\dot{w}_3\neq 0$ and/or $Q_3\neq 0$. Once again, our point of view is to test and constrain possible models of interacting dynamical DE, assuming that $\Lambda$CDM  is a model which well describes, ``only'' in an effective way,  the observations on the Universe so far.

The observable $K_0$ is written in terms of measurable variables, through \eqref{49}. A precise estimate of $K_0$ is highly nontrivial, and goes beyond the scope of our paper.
There are in fact two kind of difficulties in evaluating $K_0$. The first is that, at the moment, the quantities in terms of which $K_0$ is expressed (the Hubble constant $H_0$, the matter EoS parameter $\Omega_1$, and the kinematic variables $q(t)$ and $j(t)$, all evaluated at the present epoch) are known with large errors, in particular the jerk $j_0$, not to mention the known existing tension on the value of $H_0$. The situation will improve drastically in the next future, since, for instance, one of the aims of the forthcoming Euclid experiment is to refine the measure of the kinematic variables, which therefore will be known with much greater accuracy. Under this respect, $K_0$ is a variable which we believe will become very interesting in the future. 
\\
The other difficulty concerns the type of analysis which should be performed.  In fact, particular care should be payed in Cosmology when dealing with ``experimental'' data, which should be treated according to the Bayesian analysis. We are not experts in this kind of analysis, and, even if the observables in terms of which $K_0$ is expressed were known with smaller errors, we prefer to leave this task to professionals, whenever the kinematical variables will be known with more precision. 
\\
Therefore, although a precise Bayesian analysis to determine $K_0$ is premature, and although our paper focuses on the formal aspects of a theory of interacting dynamical Dark Energy,  in the final part of this Section we give a preliminary, albeit rough, estimate of $K_0$ based on the publicly available data sets.
\\
Concerning the deceleration $q_0$ and the jerk $j_0$, evaluated at our epoch, we report in Table 2 four maximum likelihood values, with their 68\% confidence intervals:
\\
\begin{center}
\begin{tabular}{|c|c|c|c|c|}
\hline 
& $a$ & $b$ & $c$ & $d$\\
\hline
$q_{0}$  & $-0.644 \pm 0.223$ & $-0.6401 \pm 0.187$  & $-0.930 \pm 0.218$ &$ -1.2037\pm 0.175$ \\
$j_{0}$     & $1.961\pm0.926$  & $1.946\pm0.871$ & $3.369\pm1.270 $  &$5.423\pm1.497$ \\
\hline
\end{tabular}
\\
\vspace{.2cm}{\footnotesize {\bf Table 2.}
Deceleration and jerk.}
\end{center}
The observational constraints for the deceleration parameter $q_0$ and the jerk $j_0$ reported in Table 2 were recently obtained in \cite{Capozziello:2018jya}, and $a$, $b$, $c$ and $d$ refer to the following different combinations of low redshift datasets:
\begin{description}
\item[$a$:]
BAO +Masers+TDSL+Pantheon, \\ where BAO stands for the observations from Baryon-Acoustic-Oscillations  \cite{Beutler:2011hx,Beutler:2012px,Blake:2012pj,Anderson:2012sa,Anderson:2013zyy}, Masers is the Megamaser Cosmology Project \cite{Evslin:2017qdn,Gao:2015tqd,Kuo:2012hg,Reid:2012hm}, TDSL means time-delay in strong lensing measurements by H0LiCOW experiment \cite{Bonvin:2016crt} and Pantheon are the data for SNIa in terms of $E(z)$ \cite{Gomez-Valent:2018hwc,Riess:2017lxs}
\item[$b$:]
 $a$ + $H_0$ measurement done in \cite{Riess:2016jrr,Riess:2018byc}
\item[$c$:]
$a$ + $H(z)$ measurements (Hubble parameter data (OHD) as a function of redshift \cite{Pinho:2018unz})
\item[$d$:]
all the data ($a$ + $H_0$   + $H(z)$).
\end{description}

According to the $d$-dataset in Table 2, which contains all the others, it is apparent that the $\Lambda$CDM value $j_0=1$ is incompatible with data, at 3.06$\sigma$ confidence limit.
\\
Concerning $K_0$, it is convenient to consider the dimensionless quantity $K_0/H_0$, in order to get rid of the well known tension existing on the Hubble constant \cite{Riess:2016jrr,Riess:2018byc,Ade:2015xua,Ade:2015rim}. A rough and preliminary estimate, which takes into account the values of $q_0$ and $j_0$ given above and the value of the matter density parameter $\Omega_1$, which, according to the latest SN Ia measurements from the Pantheon Catalogue \cite{Scolnic:2017caz}, is 
\begin{equation}
\Omega_1=0.298\pm0.022\ ,
\label{50}\end{equation}
gives the four values listed in Table 3:
\begin{center}
\begin{tabular}{|c|c|c|c|c|}
\hline
& $a$ & $b$ & $c$ & $d$\\
\hline
$K_{0}$  & $-0.516 \pm 0.685$ & $-0.512 \pm 0.622$  & $-1.026 \pm 0.931$ &$ -1.817\pm 1.091$ \\
\hline
\end{tabular}
\\
\vspace{.2cm}{\footnotesize {\bf Table 3.}
Estimates of $K_0.$}
\end{center}
All the above values of $K_0$ are compatible with the $\Lambda$CDM value $K_0=0$
within 1 to 2$\sigma$. Therefore, according to the data available so far, there is no evidence against  $\Lambda$CDM model. A more accurate, constrained analysis might be done following the Bayesian methods in cosmology \cite{Trotta:2017wnx}, but for the moment our aim is just to give an estimate of the right hand side of our result \eqref{34}, by means of observable quantities.

\section{Conclusions}

The points where the $\Lambda$CDM model creaks are more and more. An example of these weaknesses is the well known tension on the measurements of the Hubble constants $H_0$. The value given by the Planck collaboration \cite{Ade:2015xua,Ade:2015rim} in the framework of the 
$\Lambda$CDM model is incompatible with other, model independent, estimates \cite{Riess:2016jrr,Riess:2018byc}. The inconsistencies become milder if  a dynamical DE is invoked \cite{DiValentino:2017rcr}. Therefore, there are  strong motivations to investigate models of dynamical DE which, in the most general case, displays an EoS with a time dependent coefficient $w_{\mbox{\tiny DE}}(t)$ and which may interact, in principle, with matter and/or radiation through a (partial) breaking $Q_{\mbox{\tiny DE}}(t)$ of the covariant conservation of the energy momentum tensor.

The main and new result of this paper is represented by the relation \eqref{33}, which must be satisfied, at any time, by any model of interacting dynamical DE
\begin{equation}
\Omega_{\mbox{\tiny DE}}\dot{w}_{\mbox{\tiny DE}} + \frac{8\pi G}{3H^2}Q_{\mbox{\tiny DE}}w_{\mbox{\tiny DE}} = K\ ,
\label{51}\end{equation}
where $K(t)$ is expressed in terms of measurable quantities. The above equation is a differential equation for the DE EoS parameter $w_{\mbox{\tiny DE}}(t)$ with time dependent coefficients, one of which is the interaction $Q_{\mbox{\tiny DE}}(t)$. It must hold at any time, in particular must be satisfied by phantom DE models, crossing the $\Lambda$CDM point $w=-1$ in both directions.
The equation \eqref{51} is a constraint on the possible parametrization of Dark Energy which, in its most general dynamical form, turns out to depend on its interactions, with Dark Matter in particular. It is not surprising that it must be so, but the explicit form of how this mutual dependence is realised was not known so far. We proposed a new observable, which we called $K_0$, which measures the relation between Dark Energy and its interactions.
The analysis which led to \eqref{51} is model independent, in the sense that we only assumed that the scale factor $a(t)$ appearing in the Robertson-Walker metric \eqref{4.1} obeys the Friedmann equation {\eqref{11}} and that the quantities involved in $K(t)$ are the density parameters and the kinematic variables, hence are directly measurable.

An important consequence of \eqref{51} is that, according to the $\Lambda$CDM model,  it must hold
\begin{equation}
\Lambda\mbox{CDM}\ \ \Rightarrow\ \  K_0=0\ .
\label{52}\end{equation}
Any deviation from this value must be interpreted as a failure of the $\Lambda$CDM model.  Low-redshift data show that, at present time,  $j_0\neq 1$ \cite{Capozziello:2018jya}, which seems to indicate a failure of $\Lambda$CDM. It would be greatly interesting to give an accurate estimate of $K_0$, according to the available observational data, but this task goes beyond the scope of this paper, also because the available observational data, expecially for what concerns the kinematical variables $q(t)$ \eqref{6} and $j(t)$ \eqref{7}, are affected by large errors \cite{Visser:2003vq,Capozziello:2018jya}, which make difficult any decisive claim within 3$\sigma$. Hopefully, the Euclid space mission, whose launch date is expected in 2021, will drastically improve the experimental situation. With this $caveat$, we gave a preliminary and rough estimate of $K_0$, which is compatible, within $3\sigma$, with zero, hence with the $\Lambda$CDM model. But, again, a much more accurate evaluation will be possible in the future.

Finally, the relation \eqref{51} can be read in several ways, depending whether the DE is interacting or not ($Q_{\mbox{\tiny DE}}\neq 0$ or $Q_{\mbox{\tiny DE}}=0$). It is important to emphasize this point because previous attempts to get informations on the Dark Sector rely on particular assumptions. Our result may provide a model independent description of the Dark Sector, as well as a constraint for generic parametrizations of the EoS coefficients $w_{\mbox{\tiny DE}}$ and of the 
interactions $Q_{\mbox{\tiny DE}} (t)$.

\vspace{1cm}

%
%


\end{document}